\documentclass[nologo,url,11pt,a4paper]{ETHpaper}

\usepackage{balance}  
\usepackage{graphics} 
\usepackage{times}    
\usepackage{url}      
\usepackage{multirow} 
\usepackage{rotating} 
\usepackage{subfigure}
\usepackage{epstopdf}
\usepackage{epsfig}
\usepackage{amsmath}
\usepackage{float}
\usepackage[titletoc,toc,title]{appendix}

\begin{document}

\title{Who Watches (and Shares) What on YouTube? And When? \\Using Twitter to 
Understand YouTube Viewership}

\titlealternative{Who Watches (and Shares) What on YouTube? And When? Using 
Twitter to Understand YouTube Viewership\\
\textbf{In Proceedings of the 7th International ACM Conference on Web Science 
and Data Mining 2014}}

\author{Adiya Abisheva, Venkata Rama Kiran Garimella, David Garcia, Ingmar 
Weber}

\maketitle

\begin{abstract}
By combining multiple social media datasets, it is possible to gain insight into 
each dataset that goes beyond what could be obtained with either individually. 
In this paper we combine user-centric data from Twitter with video-centric data 
from YouTube to build a rich picture of who watches and shares what on YouTube. 
We study 87K Twitter users, 5.6 million YouTube videos and 15 million video 
sharing events from user-, video- and sharing-event-centric perspectives. We 
show that features of Twitter users correlate with YouTube features and 
sharing-related features. For example, urban users are quicker to share than 
rural users. We find a superlinear relationship between initial Twitter shares 
and the final amounts of views. We discover that Twitter activity metrics play 
more role in video popularity than mere amount of followers. We also reveal the 
existence of correlated behavior concerning the time between video creation
and sharing within certain timescales, showing the time onset for a coherent 
response, and the time limit after which collective responses are extremely 
unlikely. Response times depend on the category of the video, suggesting Twitter 
video sharing is highly dependent on the video content. To the best of our 
knowledge, this is the first large-scale study combining YouTube and Twitter 
data, and it reveals novel, detailed insights into who watches (and shares) what 
on YouTube, and when.
\end{abstract}

\section{Introduction}\label{sec:introduction}

On July 11, 2013, @justinbieber tweeted: ``so many
activities it is making my head spin! haha \url{http://t.co/Gdg615ZZGX}'', 
sharing a link to a short YouTube movie clip. In one day, the  video received 
more than 100,000 views, and its owner commented: ``So I checked my email today
to find 500 new mail... WTF I thought.... 5 mins later I discover that
Justin Bieber has tweeted this video...''. The viewers of that video
came from the 40 million followers that Justin Bieber has in Twitter, including 
large amounts of pop-loving teenagers that retweeted the video link more than 
800 times in the following days.

The above example illustrates how the combination of Twitter and YouTube data 
provide insights on \emph{who watches what on YouTube and when}. In this 
article we combine large datasets from both online communities, aiming at a 
descriptive analysis of the demographics and behavioral features of YouTube 
viewership through Twitter video shares. In our analysis, the \emph{``who''} 
refers to the identity of Twitter users, as displayed on their public profile. 
We quantify this identity in three facets: a) demographic variables such as 
gender and location, b) social metrics that include reputation and impact 
metrics in the Twitter follower network, and c) personal interests and political
alignment inferred from profile descriptions and followed accounts. The 
\emph{``what''} refers to features of the videos, including i) the YouTube 
category, and ii) the popularity of videos in terms of views or likes.  The
\emph{``when''} is the time lapsed between the creation of a video and its 
sharing in Twitter, measuring the time component of individual and collective 
reaction patterns to YouTube videos.

With a combined dataset of Twitter data and YouTube videos we can answer 
questions about the interaction between both communities. First, we explore the 
purpose of social sharing, distinguishing regular and promotional Twitter 
accounts linked to a particular YouTube channel. We then analyze to which extent 
the content of the videos watched by a user is similar to their interests on 
Twitter. Using features extracted from Twitter, we are able to quantify factors 
such as social sharing and influence and infer their effect on the videos 
consumed on YouTube. We also look at the role of political alignment in YouTube 
video sharing, comparing the shared topics and reaction patterns of
individuals depending on their political activity on Twitter.

We analyze the times between video creation and social sharing, looking for 
factors that mediate the speed of video sharing. We find the demographics of 
users that share videos earlier than the rest, and compare how different 
categories elicit faster or slower reactions in Twitter. Finally, we explore the
relation between the early Twitter shares of a video and its final popularity. 
To do so, we designed a model that includes social impact and reputation metrics 
of the early watchers of the video, providing early forecasts of a video's 
ultimate popularity.

\section{Related Work}\label{sec:relatedwork}
Since we answer ``who?'', ``what?'' and ``when?'', we describe related work done 
on Twitter profiles and online demographics, YouTube viewership and content and 
temporal behaviour patterns.

\subsection{Online Demographics}
Related work on ``who'' \emph{does} ``what'' in Web search has been done in 
Weber and Jaimes \cite{weberjaimes11wsdm} where authors analyze query logs of 
2.3 million users form a web search US engine. Even though our work performs 
analysis on Twitter and YouTube users rather than Web search users, methodology 
used in previous study is of high relevance for our research. More closely 
related work on Twitter demographics was performed in Mislove et al. 
\cite{mislove2011} where authors investigate whether Twitter users are a 
representative sample of society. By using (optionally) self-reported and 
publicly visible data of Twitter users, authors compared demographics of Twitter 
US users to the US population along three axes. On the geographical dimension, 
findings showed that Twitter users are overrepresented in highly populated US 
counties and underrepresented in sparsely populated regions due to different 
patterns of adoption of social media across regions. Across gender, the male 
Twitter population is greater than female especially among early Twitter 
adopters, but male bias decreases as Twitter evolves. On race/ethnicity authors 
show the distribution is highly geographically-dependent. Another study on 
demographics by Goel et al. \cite{GoelHS12} shows that user demographics (age, 
gender, race, income etc.) can be inferred from Web browsing histories. Finally, 
Kulshrestha et al. \cite{KulshresthaKNG12} investigate the role of offline 
geography in Twitter and conclude that it has a significant role in social 
interactions on Twitter with more tweets and links exchanged across national 
boundaries.

\subsection{Research on Twitter Data}
Apart from demographics, Twitter has also been studied from other perspectives: 
prediction of trends/hashtags \cite{Asur2011, Wang2012, Kairam2013}; notions of 
influence in Twitter \cite{ChaHBG10} and using Twitter predictive data for 
elections and discovering political alignment of users \cite{Conover2012, 
Conover2011}. Asur et al. \cite{Asur2011} studied trending topics/hashtags and  
discovered that the content of a tweet and retweeting activity rather than user 
attributes such as influence, number of followers and frequency of posting are 
the main drivers for spotting the trend and keeping it alive. In Wang and 
Huberman \cite{Wang2012} a model for attention growth and persistence of trends 
is presented and is validated on trending topics in Twitter. In another work on 
trending topics, hashtags in Twitter can be clustered according to the temporal 
usage patterns of the hashtag: before, during and after peak of its popularity. 
Furthermore, the class of the hashtag correlates with social semantics of 
content associated with the hashtag (Lehmann et al.\ \cite{Lehmann2012}). In a 
study on differences of search activity of trending topics in the Web and 
Twitter, Kairam et al. \cite{Kairam2013} reveal that information-seeking and 
information-sharing activity around trending events follows similar temporal
dynamics, but social media leads Web search activity by 4.3 hours on average. 
More generalized study on differences between Web and Twitter search by Teevan 
et al. \cite{Teevan2011} found that timely and social information are primary 
drivers for searching on Twitter, compared to more navigational search on the 
Web; Twitter search is more used to monitor new content, while search on the Web 
is performed for developing and learning about a topic. Another perspective is 
the notion of influence in Twitter using followers, retweets and mentions 
studied by Cha et al.\ \cite{ChaHBG10} with main finding that having a lot of 
followers does not necessarily mean having a high influence. Another line of 
work uses Twitter to monitor political opinions, increase political 
mobilization, and possibly predict elections' results. Conover et al. 
\cite{Conover2011} present several methods to discover political alignment of 
Twitter users by analysing the network of political retweets and hashtags usage. 
In subsequent work \cite{Conover2012}, authors go beyond discovering political 
groups in Twitter, and analyse interaction dynamics of politically aligned 
subcommunities. Their findings show that right-leaning Twitter users produce 
more political content, spend a greater proportion of their time for political 
conversation and have more tightly interconnected social structure which 
leverages broad and fast spread of information.

\subsection{YouTube Video Consumption}
Ulges et al. \cite{UlgesBK13} use YouTube concepts to predict demographic 
profile of viewers and also try to use demographics estimated from views 
statistics to predict the concepts of a video. They show that the use of 
demographic features improves the quality of prediction. Concerning YouTube 
video views, researchers have analyzed time series \cite{Cranesor}, and 
predicted the final views count based on properties of growth on YouTube 
\cite{Szabo2010}. For instance, Crane and Sornette \cite{Cranesor} perform 
analysis of collective responses to YouTube videos through their time series of 
views. Among the classes, the most usual were videos that have a fast decaying 
amount of views, receiving negligible amounts of views soon after their 
creation. Laine et al.\ \cite{Laine2011} highlighted the role of exogenous 
factors (such as interest groups) in the activity of YouTube viewers and Qiu et 
al.\ \cite{Qiu2012} suggested two different mechanisms that drive YouTube 
viewership: popularity and quality filtering. On popularity of videos in 
YouTube, Figueiredo et al. \cite{Figueiredo2011} found copyright videos gain 
90\% of their views early in lifetime compared to top listed YouTube or randomly 
chosen videos; top listed videos show \emph{quality} popularity dynamics pattern 
opposed to copyright and random videos exhibiting \emph{viral}, word-of-mouth, 
dynamics. And finally, related videos and internal search are most contributing 
towards content dissemination, but for random videos social link is also a key 
factor. On politics in YouTube, recent work by Garcia et al.\ \cite{Garcia2012a} 
performs analysis of collective responses to the YouTube videos of US political 
campaigns and reveals differences in collective dynamics that suggest stronger 
interaction among right-leaning users. Weber et al.\ \cite{weberetal13www} use 
YouTube video tags and conclude that general YouTube videos are not polarized in 
terms of audience, but for subclasses of apolitical videos (e.g., tagged as 
``army'') an audience bias can be predicted (right-leaning in this case). 
Finally, in Crane et al.\ \cite{Crane2010} collective responses to the videos of 
Saddam Hussein's death show an extremely fast response and relevance of news and 
politics for YouTube viewers.

\subsection{On Human Behaviour}
Since our analysis involves the ``when'' dimension of video shares on Twitter, 
we  review work on temporal patterns in human behaviour. Quantitative 
understanding of human behavior, also known as human dynamics, got a new turning 
point in 2005 after work by A.-L. Barab\'{a}si \cite{barabasi05}, where author 
looked whether the timing of human activities follows any specific pattern. 
Results  showed that there are bursts of intensive activity interchanged with 
long periods of inactivity (Pareto distribution) rather than events happening at 
regular time intervals (Poisson). Since 2005 more studies on the inhomogeneous 
nature of temporal processes in human dynamics have been performed 
\cite{Crane2010, karsai12, wu10}. Various proxies were used to get timing of 
human activity, e.g., mobile records, web server logs, SMS etc. Recent study by 
Wu et al. \cite{wu10} suggests time patterns follow bimodal distribution with 
bursts of activity explained by power-law distribution in the first mode and 
exponentially distributed initiation of activity in the second mode.

\section{Data Set}\label{sec:dataset}
We collected data from Twitter and YouTube for our analysis, and related the 
datasets by looking at instances where links to videos were shared on Twitter. 
This section describes how we obtained the 87K Twitter users, 5.6 million 
YouTube videos, and 15 million video sharing events we analyzed in greater 
detail. Data sets are available at \url{
http://web.sg.ethz.ch/users/aabisheva/2013_YouTube_Twitter_ETH_QCRI/index.html}.

\subsection{Twitter}\label{sec:data:twitter}
The data acquisition starts with a 28 hour time slice from 6/6/2013 21:00 to 
8/6/2013 1:00 (AST) of all public Tweets containing any URL provided by GNIP, a 
reseller of Twitter data. Of these tweets, only tweets by users with at least 
one follower, one friend, has non-empty profile location and English as profile 
language were considered. 1,271,274 tweets containing a URL from 
\url{http://www.youtube.com} or \url{http://youtu.be} where identified. URLs 
shortened by Twitter's default URL shortener \url{t.co} were automatically 
unshortened, but other services were not considered. From this set, 200K 
distinct tweets were sampled uniformly at random. These tweets account for 
177,791 distinct users. Out of these, 100K users were sampled uniformly at 
random.

For each of these users we obtained (up to) their last 3,200 public tweets. In 
12,922 cases this failed because the user account had been removed or made 
private. Along with the tweets, we obtained the user's public profile, 
containing the user-defined location, their followers and friends count and the 
set of (up to) 5,000 friends (= other Twitter users the user follows) and 5,000 
followers (= other Twitter users who follow this user). 96.8\% of our users had 
less than 5,000 followers and 98.9\% had less than 5,000 friends. We also got 
the profile information for all these friends and followers. Finally, we had 
17,013,356 unique tweets with 5,682,627 distinct YouTube video IDs, 19,004,341 
friends and 22,182,881 followers for the 87,076 users. From this data we 
extracted a number of features related to (i) demographics, (ii) location, (iii) 
interests, and (iv) behavior on Twitter.

\textbf{Demographics.}
We used a name dictionary to infer the self-declared gender of a Twitter user 
using common first names and gender from 
\url{http://www.ssa.gov/oact/babynames/limits.html}. To detect a subset of 
potential parents, we scanned each user's ``bio'' for mother/mom/wife or 
father/dad/husband using exact token match. Similarly, we identified a subset of 
potential students by scanning the bios for student/study/studying.

\textbf{Location.}
Each of the users profile locations was run through Yahoo!'s Placemaker 
geo-coder, \url{http://developer.yahoo.com/yql/console/}, and for 61,250 
profiles, a location could be identified. For the 23,416 users with an 
identified location in the US we checked if their city matched a list of the 100 
biggest US cities from \url{http://www.city-data.com/top1.html}. This gave us an 
estimate of users from rural vs.\ urban areas.

\textbf{Interests.}
To detect interests of users, we chose to analyze the users they follow. These 
friends were then compared against directory information from 
\url{http://wefollow.com}\footnote{WeFollow is a website listing Twitter users 
for different topics along with a ``prominence score'', indicating importance of 
the user in the respective field. WeFollow's directory has been used in several 
academic studies \cite{liaoetal12chi, bostandjievetal12recsys, 
alzamaletal12icwsm, pennacchiottipopescu11kdd}}. Concretely, we obtained 
information for the classes Sports, Movies, News \& Politics, Finance, Comedy, 
Science, Non-profits, Film, Sci-Fi/Fantasy, Gaming, People, Travel, Autos, 
Music, Entertainment, Education, Howto, Pets, and Shows as described in 
Table~\ref{tab:wefollow}. In addition to the information from wefollow.com, we 
labeled 32 politicians or party accounts on Twitter as either Democrats (13) or 
Republicans (19). The same list was also used by Weber, et al. 
\cite{Weber2012,Weber2013}. Users were then labeled as left or right according 
to the distribution of users they followed (if any). Following had previously 
been shown to be a strong signal for political orientation \cite{ConoverRFGMF11, 
barbera2012, Weber2012, Weber2013}.

\textbf{Behavior.}
To quantify the activity of a user on Twitter, we extracted various features 
such as their number of tweets, the fraction of tweets that are retweets, or the 
fraction of tweets containing URLs.

Finally, we also aggregated features from all YouTube videos shared by a user 
into statistics such as the average view count or the median inter-event time 
(``lag'') between video upload and sharing. These features are described in more 
detail in the next section.

\begin{table}[b]
\centering
\small
\begin{tabular}{cc}
Category & wefollow.com Interests \\ \hline
Sports & sports, baseb., basketb., soccer, footb., cricket, nfl  \\
Movies & movies  \\
News \& Politics & economics, politics, news  \\
Finance & banking, investing, finance, entrepreneur, business  \\
Comedy & comedy, comedian  \\
Science \& Technology & tech, technology, gadgets, science, socialmedia \\
Non-profits \& Activism & non-profits, non-profit, charity, philanthropy  \\
Film \& Animation & film, animation, cartoons \\
Sci-Fi/Fantasy & scifi, sciencefiction, fantasy  \\
Gaming & games, gaming  \\
People \& Blogs &  blogger, blogs, people, celebrity  \\
Travel \& Events & travel, places  \\
Autos \& Vehicles &  automotive, autos, cars, vehicles  \\
Music &  music, dance, dancer  \\
Entertainment & entertainment  \\
Education &    academic, university, education \\
Howto \& Style &    howto,  diy,    doityourself  \\
Pets \& Animals &    animals, cats, dogs, pets  \\
Shows & tv, tvshows, media  \\
\end{tabular}
\caption{Mapping of YouTube categories (left) to wefollow.com interests (right). 
The YouTube category ``Trailers'' was not mapped. The non-YouTube category 
``Finance'' was added.}\label{tab:wefollow}
\end{table}

\subsection{YouTube Activity on Twitter}\label{sec:youtubeDataset}
Given 17,013,356 unique tweets with YouTube video IDs, we retrieved 15,211,132 
sharing events and identified 6,433,570 unique YouTube video IDs. We define 
sharing event as a tweet containing valid YouTube video ID (having category, 
Freebase topics and timestamp), thus a tweet with two video IDs is considered as 
two sharing events. A fraction of videos in initial 17 million tweets were not 
valid, thus such tweets and consequently derived sharing events were removed. 
Using the YouTube API, the following data about videos was crawled within the 
period 7/7/2013 -- 1/8/2013: title, uploader username, number of views, number 
of times video has been marked as favorite, number of raters, number of likes 
and dislikes, number of comments, video uploaded time and categories to which 
videos belong. Using the Freebase API we also crawled video topics which serve 
as deprecated video tags and are helpful for searching content on YouTube, e.g., 
``hobby'', ``music'', ``book'' and many others are examples of Freebase topics 
(\url{http://www.freebase.com/}).

The cleansing stage of data contained three parts: identify noise in data, 
introduce a filter on Twitter users with ``extreme'' behaviour and introduce a 
filter on ``legacy'' YouTube videos (see Section \ref{sec:earlyadopters}). In 
our data set, noisy data (0.53\%) are those sharing events where the tweet's 
timestamp is earlier than the video's upload timestamp. Such negative lags 
spanned from 1 second up to a couple of years. We removed all such sharing 
events which seemed to occur 1) due to updated timestamp of streamed live videos 
recorded by YouTube where the time at the end of streaming is returned as 
published timestamp by YouTube API, and 2) due to altered timestamp of 
reuploaded videos by some YouTube ``privileged'' accounts. After removing noise, 
the data reduced to 15,130,439 sharing events, 5,669,907 unique video IDs and 
87,069 user IDs.

Handling the data, we came across ``non-human'' behaviour explained by automated 
video sharing. We identify Twitter accounts and YouTube channels possibly owned 
by the same user, and label such Twitter users, as \emph{promotional} since the 
primary content of such videos is advertisement. These accounts are often in top 
1 percentile of Twitter users sorted by the number of YouTube videos shared. 
Examples of such Twitter-YouTube pairs with the number of shared videos in 
brackets are: \emph{spanish\_life} -- \emph{aspanishlife} (8,119) on real estate 
advertisement and \emph{RealHollywoodTr} -- \emph{bootcampmc} (5,315) blogging 
on fitness and health, while the mean number of shares per user was found at 174 
video shares. To remove \emph{promotional} users, we applied a filtering 
mechanism based on a) similarity between usernames in Twitter and YouTube using 
longest common substring (LCS), and/or b) amount of videos in Twitter account 
coming from one YouTube source; for details refer to supplementary material 
submitted in: \url{http://arxiv.org/abs/1312.4511}. We follow an aggressive 
approach when detecting promotional users; thus, there is a possibility of some 
regular users being labeled as promotional but not the other way round. As a 
result of filtering we split Twitter accounts into 71,920 regular 
non-promotional and 15,149 promotional accounts.

\section{Who Watches What?}\label{sec:whowatcheswhat}
In this section, we present a first analysis of who (in terms of Twitter user 
features) watches and shares what (in terms of YouTube video features). Though 
we include here user features related to the inter-event time, early video 
sharers are analyzed in Section~\ref{sec:earlyadopters}.

\subsection{Cluster Analysis}
As a first picture of who watches and shares what we present a cluster analysis 
of 26,938 non-promotional, sufficiently active users who shared at least 10 
YouTube videos and had at least 10 friends matched on Twitter through WeFollow 
(see Table~\ref{tab:wefollow}). These users were clustered into eight groups 
according to the (normalized) distribution of YouTube categories of the videos 
they shared using an agglomerative hierarchical clustering algorithm with a 
cosine similarity metric \cite{karypis02cluto}. Table~\ref{tab:clustering} shows 
the results.

\begin{table}[b]
\resizebox{\textwidth}{!}{%
\begin{tabular}{ccccccccc}
 & Cluster1 & Cluster2 & Cluster3 & Cluster4 & Cluster5 & Cluster6 & Cluster7 & 
Cluster8 \\
 & (2740) & (2327) & (2493) & (5390) & (2535) & (4052) & (3697) & (3704)\\
\hline
\multirow{5}{*}{\begin{sideways}\parbox{25mm}{Discriminating 
features}\end{sideways}} & sports & animals & non-profit & music & news/politics 
& film/animation & entertainment & travel \\
 & music & music & music & non-profit & music & education & people/blogs & music 
\\
 & entertainment & entertainment & sports & sports & comedy & music & howto & 
gaming \\
 & people/blogs & people/blogs & entertainment & education & entertainment & 
non-profit & sports & science/tech \\
 & non-profit & sports & education & animals & education & sports & music & 
autos \\
\hline
\multirow{5}{*}{\begin{sideways}\parbox{15mm}{Top profile words}\end{sideways}} 
& fan & music & music & music & music & music & life & music \\
 & music & life & life & life & life & life & music & gamer \\
 & sports & fan & fan & artist & world & fan & fan & life \\
 & life & lover & world & producer & conservative & lover & live & fan \\
 & football & writer & lover & live & people & time & justin & youtube \\
\hline
\multirow{5}{*}{\begin{sideways}\parbox{19mm}{Top features}\end{sideways}} & T 
sports$^+$ & Y animals$^+$ & Y non-profit$^+$ & Y music$^+$ & Y 
news/politics$^+$ & Y film$^+$ & Y howto$^+$ & Y gaming$^+$ \\
 & Y sports$^+$ & T animals$^+$ & T non-profit$^+$ & median lag$^+$ & Y 
comedy$^+$ & Y education$^+$  & Y people$^+$ & Y science/tech$^+$ \\
 & male$^+$ & std dev. of lag$^+$ & num. usrs rtwd$^+$ & T music$^+$ & T 
news/politics$^+$ & frac. Tw other URLs$^-$  & Y entertainment$^+$ & T 
gaming$^+$ \\
 & frac. Tw other URLs$^+$ & acnt created at$^-$ & frac. of usrs Tw rtwd$^+$ & 
mean lag$^+$ & frac. Tw other URLs$^+$ & T movies$^+$ & female$^+$ & Y shows$^+$ 
\\
 & avg. frnds of frnds$^-$ & T education$^+$ & frac. Tw Y videos$^-$ & Y 
education$^-$ & leaning republic$^+$ & Tfilm$^+$ & avg rtwt count user$^+$ & 
num. vids shared$^+$ \\
\end{tabular}}
\caption{Clusters obtained by clustering normalized YouTube categories 
distributions for each user.}\label{tab:clustering}
\end{table}

We were interested to see which differences for Twitter features are induced 
when users are grouped solely according to YouTube categories. To describe the 
clusters found, Table~\ref{tab:clustering} first lists the \emph{discriminative} 
YouTube features as output by the clustering algorithm. Below it lists the 5 
most prominent terms from the Twitter bios of users in this group. These terms, 
which were \emph{not} used to obtain the clustering, give fairly intuitive 
descriptions of the user groups. Finally, the table lists features whose average 
value differs statistically significantly (at 1\%) between the cluster and all 
27K users. These features are ranked by the absolute difference between global 
and within-cluster averages, divided by the standard deviation.

Inspecting the clusters, certain observations can be made. First, the 
discriminating YouTube categories (first block of five lines) are largely 
aligned with Twitter categories that are over-represented in the corresponding 
cluster (The ``T *'' in the bottom block of five lines). This alignment we will 
investigate more in Section~\ref{sec:correlation_analysis}. Second, there are 
certain correlations between the demographics and the YouTube categories. For 
example, Cluster 1 is focused on sports and has more male users, whereas Cluster 
7 is centered around entertainment and people/blogs and has more female users. 
Recall that the clustering was done according to YouTube categories, whereas the 
demographic information comes from Twitter, indicating the possible benefits of 
the combination. Finally, the clustering also picks up a connection to political 
orientation. Concretely, Cluster 5 contains more conservative users with an 
increased interest in news and politics (more on this in 
Section~\ref{sec:correlation_analysis}).

\subsection{Demographics}
To understand the significance of the influence of variables such as gender or 
occupation on (i) the number of views, (ii) the polarization or 
controversiality\footnote{We calculate the polarization that a YouTube video 
creates on its viewers through its amounts of likes $L_v$, dislikes $D_v$, and 
total views $V_v$, through the equation $Pol_v= \frac{L_v}{V_v^{0.849}} 
\cdot\frac{D_v}{V_v^{0.884}}$. The rationale behind this calculation is the 
rescaling of the likes and dislikes ratio based on the fact that they do not 
grow linearly with each other. The exponents correspond to the base rates of the 
logarithmically transformed amounts of views, likes and dislikes. This way we 
standardize the ratio over their nonlinear relation.}, and (iii) the lag we 
applied a so-called ``permutation test''~\cite{good06springer}, which unlike 
other tests does not make assumptions on the distribution type of the observed 
variables. To test, say, the impact of stating ``student'' in the Twitter bio on 
the number of views we first computed the average view count for all views by 
the ``student'' group and compared this with the average for the complement 
``non-student'' group. Let $\delta$ be the observed difference. Then to test the 
significance of $\delta$ we pooled all the student and non-student labeled 
observations and randomly permuted the two labels to get two groups. For these 
two groups, obtained by a label permutation, a $\delta_p$ was then computed. 
This process was repeated 10,000 times to estimate the common level of 
variability in the $\delta_p$. We then marked the $\delta$ as significant if it 
was in the bottom/top 0.5\% (or 2.5\%) of the percentiles of the $\delta_p$. In 
Table~\ref{tab:demographics}, a $^{**}$ indicates that $\delta$ was in the 
bottom/top 0.5\% and $^{*}$ indicates that it was in the bottom/top 2.5\%.

For Table~\ref{tab:correlation_twitter_youtube} we used a similar procedure to 
test the statistical significance of the Spearman rank correlation coefficient. 
Here, to establish the common level of variability we randomly permuted both 
rankings to be correlated 10,000 times and observed the distribution of the 
Spearman rank correlation coefficients. If the original, actual coefficient fell 
within the bottom/top 0.5\%/2.5\% we marked it as significant.

Table~\ref{tab:demographics} shows correlations with respect to the per-user 
median (i) number of views of shared videos, (ii) polarization/controversiality 
of shared videos and (iii) of inter-event times. One of the demographic 
differences that can be spotted is that men compared to women share less popular 
(fewer views) videos earlier (smaller lag). Some differences are hidden in this 
analysis though, as \emph{both} urban \emph{and} rural users seem to have a 
lower lag (share fast). The explanation for this apparent paradox is that users 
who have either no self-declared location or where the location is outside of 
the US have a comparatively larger lag, and that the comparison is with 
non-urban and non-rural, which mostly consist of these users, see 
Section~\ref{sec:who_shares_faster} for more details.

\subsection{Correlation Analysis}
\label{sec:correlation_analysis}
In this section we analyze the relationship between Twitter user features, such 
as the number of followers or the fraction of tweets that contain a hashtag, and 
YouTube features, such as the number of views. As a simple analysis tool we 
computed Spearman's rank correlation coefficient for each pair of features. To 
simplify the presentation, we group the Twitter features into four classes. 
First, to see how ``social'' a user is we look at (i) the number of friends, and 
(ii) the number of distinct users mentioned. Second, to see how common 
``sharing'' is for a user we included the fraction of tweets that (i) are 
retweets, (ii) contain a hashtag, (iii) contain a YouTube URL, and (iv) contain 
a non-YouTube URL. Finally, we look at notions of ``influence'' that includes 
(i) the number of Twitter followers, (ii) the fraction of a user's tweets that 
are retweeted, (iii) the average retweet count of tweets that obtained at least 
one retweet, and (iv) the average number of followers of a user's followers.

For YouTube we consider the medians of (i) the number of views of videos shared 
by a user, (ii) the polarization of these videos, and (iii) the time lag of the 
video sharing events of the user. The number of comments received by videos 
shared by a user behaved qualitatively identical to the number of views and is 
omitted.

Our results are presented in Table~\ref{tab:correlation_twitter_youtube}. Each 
cell in the table links a Twitter user feature group (row) with a particular 
YouTube video feature (columns). The three symbols in the cell indicate ``+'' = 
significant (at 1\% using a permutation test as previously described) and 
positive, ``-'' = significant and negative, and ``0'' = not significant or below 
0.05. The symbols are in the order of the features listed above in the text.

Certain general observations can be made. For example, all of our notions of 
``social'' correlate with a drop in lag time, and out of the topics considered, 
News \& Politics is the one that is most consistently linked with users who 
actively share. But other observations are more complex and, for example, only 
some but not other notions of influence correlate positively with a large number 
of views.

\begin{table}[t]
\centering
\begin{tabular}{ccccccccc}
 & male & fe- & urban & rural & stu- & mo- & fa- & US \\
 & & male & & & dent & ther & ther& \\
\hline
views & 0 & $+^{**}$ & $-^*$ & 0 & 0 & $-^*$ & $-^*$ & $-^*$ \\
polariz. & 0 & $-^*$ & 0 & $-^*$ & $-^*$ & $-^*$ & $-^{**}$ & 0 \\
lag & $-^{**}$ & $+^*$ & $-^*$ & $-^*$ & 0 & $+^{**}$ & $-^*$ & $-^*$ \\
\end{tabular}
\caption{Demographics. A + indicates a positive deviation from the general 
population, - negative and 0 not statistically significant. $^{**}$ indicates 
that the significance was based on $\delta$ being in the bottom/top 0.5\%, 
$^{*}$ for the bottom/top 2.5\%.}\label{tab:demographics}
\end{table}

\begin{table}[t]
\centering
\begin{tabular}{c|cccccc}
 & views & polariz. & lag & Music & Sports & News \\
\hline
Social & - - & - - & - - & + 0 & 0 + & + + \\
Sharing & 0 - \textbf{-} \textbf{-} & 0 \textbf{-} 0 \textbf{-} & \textbf{-} 
\textbf{-} - \textbf{-} & 0 - 0 \textbf{-} & 0 + \textbf{+} 0 & + + \textbf{+} 
\textbf{+} \\
Influence & - \textbf{-} \textbf{+} 0 & - \textbf{-} \textbf{+} 0 & 0 - - 
\textbf{+} & + 0 + 0 & 0 0 + 0 & + + - + \\
\end{tabular}
\caption{Columns 1-3 show the relation between Twitter and per-user aggregated 
YouTube features. Columns 4-6 show the relation between Twitter and fractions of 
categories of YouTube videos shared for three example categories. Twitter 
features are grouped into three classes. Symbols indicate strength and direction 
of significance. Bold symbols indicate an absolute value of Spearman's Rank 
correlation coefficient > 0.1. See text for 
details.}\label{tab:correlation_twitter_youtube}
\end{table}

We also looked at relation between the Twitter user features and the fraction of 
video shares for various YouTube categories. 
Table~\ref{tab:correlation_twitter_youtube} shows results for the three example 
categories Music, Sports and News \& Politics. Again, different patterns for 
different definitions of ``influence'' can be observed. Out of the three topics, 
News \& Politics is the one that correlates most with social and with sharing 
behavior.

\subsection{Interests on Twitter vs.\ YouTube}
Given that our analysis links Twitter behavior to YouTube sharing events it is 
interesting to understand if the interests on the two platforms are aligned. 
Though we cannot reason about YouTube views not corresponding to Twitter sharing 
events, we compared the topical categories of a user's shared videos with the 
topical categories of their Twitter friends. To infer the latter, we used the 
WeFollow data described in Section~\ref{sec:data:twitter} where entries in 
WeFollow were also weighted according to their prominence score. This way, a 
user following @espn (prominence 99) is given a higher weight for sports than a 
user following @hoyarowing (prominence 23). To compare if a user's YouTube 
category distribution and Twitter friends WeFollow distributions are similar, we 
decided \emph{not} to compare these directly due to the following expected bias. 
The coverage by WeFollow for the different categories is likely to differ. For a 
popular topic such as music, the coverage is potentially over-proportionally 
good compared to less popular ones. To correct for this, we first normalize as 
follows.

Let $c_{ij}^T$ the prominence-weighted fraction of a user $i$'s Twitter friends 
that are recognized in the WeFollow category $j$. Similarly, define $c_{ij}^Y$ 
for their shared YouTube category distribution. Now normalize both of these 
matrices for a fixed category $j$ such that $\hat{c}_{ij}^T = c_{ij}^T / \sum_k 
c_{kj}^T$. This, effectively, compares users according to their relative 
interest in a given topic. This is then further normalized to obtain per-user 
probability distributions via $\tilde{c}_{ij}^T = \hat{c}_{ij}^T / \sum_k 
\hat{c}_{ik}^T$, similarly for $c_{ij}^Y$.

Then, for each category $j$, we look at the distribution of the differences 
$\tilde{c}_{ij}^T - \tilde{c}_{ij}^Y $ across users $i$. Categories where this 
difference is positive indicate a relatively higher importance/preference for 
Twitter, cases with a negative preference indicate a relatively higher 
importance for YouTube. Generally, the differences were very small with the 
median difference not exceeding .04 in absolute value for any category and being 
smaller than .01 for more than half. Some categories such as Film \& Animation 
were very slightly more prominent on YouTube (indicated by the negative mean and 
median), whereas Science \& Tech was slightly more prominent on Twitter. This 
analysis was done for active users with at least 10 shared videos and at least 
10 friends matched on WeFollow.

\subsection{Politics in Twitter and YouTube}\label{sec:politics}
To see how \emph{Politics} is introduced in both Twitter and YouTube, we had the 
following questions in mind: a) which political user groups share more 
politically charged content, b) what is the most frequent content of each 
political user group.

As mentioned in Section~\ref{sec:data:twitter}, to separate users into political 
groups we followed a US bipartite system with audience divided into left (L) and 
right (R) users. Users that followed more of the 13 left seed users were marked 
as left-leaning, users that followed more of the 19 right seed users were marked 
as right-leaning and users with a split preference or not following any seed 
user were marked as apolitical. Our approach resulted in three disjoint sets of 
left users $U_L$ ($|U_L| = 11,217$), right $U_R$ ($|U_R| = 1,046$) and 
apolitical users $U_A$ ($|U_A| = 57,672$).

We addressed question a) by looking at how much L, R, A users share videos in 
the category \emph{News \& Politics}. If left-leaning user $u_L$ shared set of 
videos $V_L$ with a subset of videos in the category \emph{News \& Politics}, 
$V^{\text{\emph{News\&Politics}}}_L \in V_L$; then we looked at the distribution 
of ratio of number of political video shares to total amount of shares per each 
$u_L$, $u_R$ and $u_A$: $r_{\{u_L,u_R,u_A\}}=$
$\frac{|V^{\text{\emph{News\&Politics}}}_{\{L,R,A\}}|}{|V_{\{L,R,A\}}|}$. On 
average mean ratio of videos with political content for each user population is: 
$\mu_L = 0.06$, $\mu_R = 0.29$, $\mu_A = 0.05$, which confirms right users share 
more news and politics related videos compared to left users and apolitical 
users.

To answer question b) we calculated topic distributions of videos per each 
political user category and rank topics in each user group according to their 
frequency. In order to statistically compare the ranking of topics across 
groups, we applied the distance between ranks of topics method by Havlin 
\cite{havlin1995}. If $R_1(\lambda)$ is the rank of topic $\lambda$ in user 
group 1 and $R_2(\lambda)$ is the rank of the same topic $\lambda$ in user group 
2, distance $r_{12}(\lambda)$ \emph{between the ranks of topic $\lambda$ in two 
user groups} is $r_{12}(\lambda) = |R_1(\lambda) - R_2(\lambda)|$. Thus, the 
distance \emph{between two user groups} is defined as the mean square root 
distance between the ranks of all common topics: $r_{12} = 
(\frac{1}{N}\sum_{\lambda}r^2_{12}(\lambda))^{\frac{1}{2}}$, where $N$ is the 
number of common topics across user groups. We summarize the distance metric 
across four user groups: Left, Right, Apolitical and all population (Left, Right 
and Apolitical) with $N=23,844$ and $R_{
\text{\emph{max}}} = 281,265$ in Table~\ref{tab:distanceTopicsRankingPol}.

\begin{table}[t]
\centering
\begin{tabular}{c|c|c|c|c}
& Left & Right & Apolitical & All \\ \hline
Left & - & 35733.87 & 33807.2 & 25722.16 \\
Right & 35733.87 & - & 49314.69 & 37913.44 \\
Apolitical & 33807.2 & 49314.69 & - & 23879.92 \\
All & 25722.16 & 37913.44 & 23879.92 & - \\
\end{tabular}
\caption{Distance across political, apolitical and all user groups.}
\label{tab:distanceTopicsRankingPol}
\end{table}

We find that the distances from right users is maximum to left, apolitical and 
all, and left and apolitical are close to each other in terms of distance. This 
suggests that right users have their own hierarchy of topics distinguished from 
left and apolitical users, while latter groups have more similar topics. To 
support our findings in distance between topic ranks, we look at the most 20 
frequent Freebase topics for each user group. Right users share more politically 
charged content including politicians (Barack Obama, Alex Jones, Ron Paul), news 
channels (Russia Today, The Young Turks), military-related keywords (Gun, 
Police) and concepts (USA). Conversely, left-leaning users have similar 
interests as apolitical, giving priority to entertainment videos. For example, 
``Barack Obama'' topic (Freebase ID \emph{/m/02mjmr}) is placed 30$\text{th}$ 
popular among left users and 1$\text{st}$ among right population.

Results of a) and b) support each other and give the following picture on 
political engagement of L/R/A user groups. For left users, a) says they act as 
apolitical users and on average do not share much political videos, with b) 
confirming that among top 20 video topics of left users none relate to politics. 
And for right users, a) states that they share more political content which is 
supported by b) where 9 out of top 20 topics have government, news, politics 
related concepts.

\label{sec:obama}
A possible explanation of the fact that the supposed left is much closer to the 
apolitical set than the right is that following \emph{@barackobama} is not a 
good proxy for political orientation due to his popularity in social media. To 
show that following \emph{@barackobama} \emph{is} a signal for both a) being 
more politicized and b) being more left-leaning we perform a number of 
statistical tests on differences between \emph{@barackobama} followers and 
non-followers. For a) we count the number of known political hashtags such as 
\emph{\#p2, \#tcot, \#obama, \#ows} and others for both user groups. For b) we 
count the number of words \emph{``liberal'', ``progressive'', ``democrat''} and 
\emph{``conservative'', ``republican''} in the bios of both followers and 
non-followers. The idea here is that the first (abbrev. L-words) and second 
(abbrev. R-words) word groups are indicators of someone being left- and right- 
aligned respectively. Table \ref{tab:obamaCelebrity} shows results with a clear
message: followers of \emph{@barackobama} are at least 4 times more likely to be 
left-aligned compared to non-followers (0.70\% vs. 0.16\%) and are twice more 
likely to insert political hashtags in their tweets compared to non-followers 
(20.5\% vs. 10.3\%) . Ratios were tested with a Chi-square test for equality of 
proportions with a 95\% confidence interval with  significance at $p$-value 
$<10^{-15}$.

\begin{table}[b]
\centering
\begin{tabular}{c|c||c|c||c}
& Political \# & L-words & R-words & Total \\ \hline
followers & 20.5\% (3829) & 0.70\% (130) & 0.28\% (53) & 18664 \\
$\neg$followers & 10.3\% (8615) & 0.16\% (131) & 0.35\% (281) & 83789 \\
\end{tabular}
\caption{Percentage and counts in brackets of users having political tweet 
hashtags, ``left''- and ``right''- words in account description of 
\emph{@barackobama} followers and non-followers.}
\label{tab:obamaCelebrity}
\end{table}

\section{Early Video Adopter}\label{sec:earlyadopters}
This section answers a) who shares video content faster and b) which information 
is shared faster. Thus, we look at another dimension linking Twitter and YouTube 
-- the time lag between the video upload and the sharing event on Twitter, also 
known as inter-event time or lag and denoted as $\Delta t$. We perform 
inter-event time analysis on a system level and per user. For system inter-event 
time analysis we collected time lags, $\Delta t^{w}_{v}$, per sharing event 
(tweet $w$, video $v$), resulting in time lag collection $T$, i.e., $\forall w 
\in \text{\emph{TWEETS}}, \forall v \in \text{\emph{VIDEOS}}, \Delta t^{w}_{v} 
\in T$, where \emph{TWEETS} is a set of all tweets in data set and \emph{VIDEOS} 
is a set of all videos. Thus, a user having more than one tweet with video has 
more than one time lag; similarly, a video that has been shared more than once 
will have more than one time lag; thus, several sharing events of a video are 
considered as separate sharing events, and time lag of each such event becomes
a member of collection $T$. For per user inter-event time analysis we calculated 
median time lag per each user $u$, 
$\langle \Delta t \rangle_u^{\text{\emph{median}}}$.

One limitation of the YouTube dataset was a non-uniform distribution of video 
age. Thus, we removed videos before certain epochs when YouTube and Twitter 
underwent changes. First, Twitter was founded in 2006, nearly one year after 
YouTube, thus we cannot sensibly study sharing of videos uploaded in 2005-2006. 
The next disrupting event is the introduction of Twitter share button in YouTube 
on 12/8/2010, changing the ease of sharing. Additionally, our crawled dataset 
had another constraint: a limit of 3,200 tweets per user which mainly has effect 
on tweets sample of active Twitter users. Selected sample potentially contains 
only recent tweets and thus relatively ``young'' videos in those tweets. In 
order to get a uniform age of shared videos, we determined a cutoff time for 
discarding videos of certain age at which amount of shares per user is affected 
the least. We removed videos older than $\theta = $ 1/1/2012, which 
automatically discards tweets containing such videos. The filtered data set 
contained 11,697,550 sharing events for 2,510,653 distinct videos coming from 
70,874 non-promotional users.

\subsection{Who Shares Faster in Twitter}
\label{sec:who_shares_faster}
\begin{figure}[t]
\centering
 \epsfig{file=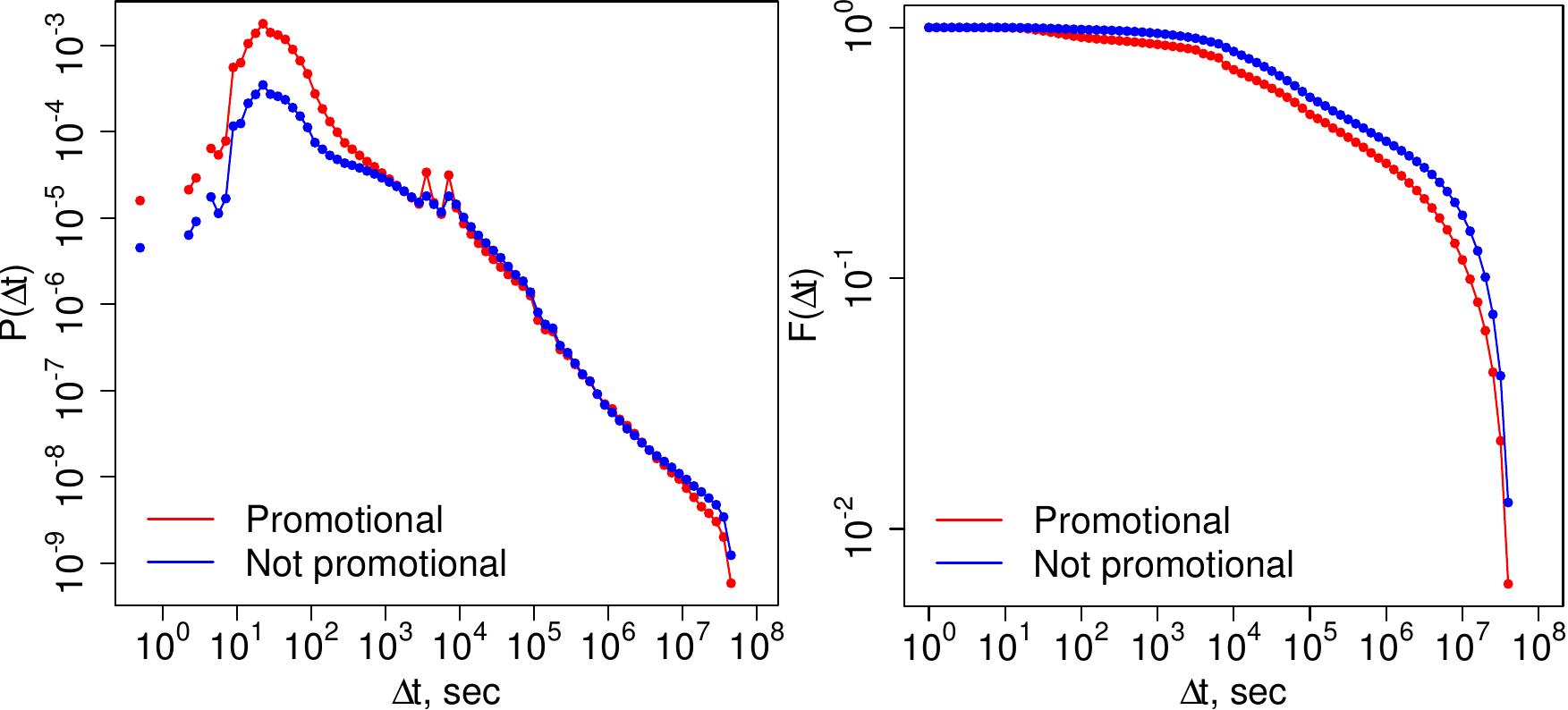, width=\textwidth}
 \caption{Inter-event time distribution $P(\Delta t)$ and accumulative time 
distribution $F(\Delta t)$s of promotional (red) and non-promotional (blue) 
accounts.}
 \label{fig:promoNotPromoTime}
\end{figure}

Question a) was addressed by comparing inter-event times per different user 
groups. We first looked at time differences between promotional and 
non-promotional Twitter accounts, see Figure~\ref{fig:promoNotPromoTime} from a 
system's point of view (rather than aggregating per-user). Visually, we observe 
that promotional accounts are faster at sharing content compared to regular 
users, see the head at $P(\Delta t)$ and tail at $F(\Delta t)$. Statistically, 
median(promo)$ = 10^{4.8}$sec (18 hours), median(non-promo) $ = 10^{5.1}$sec (38 
hours). Within an hour promotional accounts have twice amount of shares compared 
to non-promotional accounts which constitutes twentieth and tenth percentile 
respectively.

Having confirmed that there is a difference between human and ``machine'' 
behaviour, we performed a per-user inter-event time analysis for different user 
groups of non-promotional accounts. For each user group $U_G$ we calculate the 
median lag per group (median of users' medians): $\Delta t_{G} = \langle \langle 
\Delta t \rangle_u^{\text{\emph{median}}} \rangle_{u \in 
U_{G}}^{\text{\emph{median}}}$.

For example, in Section~\ref{sec:politics} we looked at who shares what per 
political user groups (Left vs.\ Right). Here we find that on average right 
users share newly uploaded video content at least 3 days earlier compared to 
left users. Note that the set of videos being shared is different though. Our 
findings on the median of the median inter-event times for various user groups 
are presented in Table~\ref{tab:interevent_times_groups}. Time differences in 
the per topic medians follow the same trend as the overall distribution (not 
presented here), so the observed differences cannot solely be explained by 
differences in category preferences for different user groups.

We highlight the following observations on who shares faster: concerning 
location, urban users are around 14 hours faster than rural users, and across 
gender women are much slower compared to men. Globally, people from Indonesia 
and Thailand have a reaction time in the order of a day, where as the greatest 
lag in the order of a half of a month is observed from people tweeting in 
Brazil. But as we selected only English profiles the results for other countries 
might be conflated with other factors.

\begin{table}[b]
\centering
\begin{tabular}{ccc}
Category & Med. int. time & num. users \\
\hline
promotional & 27 & 15132 \\
non-promotional & 141 & 70874 \\
\hline
promotional urban & 40 & 2096 \\
promotional rural & 25 & 1693 \\
\hline
non-promotional urban & 143 & 5951 \\
non-promotional rural & 157 & 5928 \\
\hline
left & 163 & 11356 \\
right & 90 & 1355 \\
\hline
male & 142 & 24263 \\
female & 187 & 16293 \\
\hline
student & 156 & 877 \\
not student & 141 & 69997 \\
\hline
mother & 191 & 450 \\
not mother & 141 & 70424 \\
\hline
father & 85 & 356 \\
not father & 141 & 70518 \\
\end{tabular}
\caption{Comparison of median of median inter-event times (in hours) for various 
groups of users}\label{tab:interevent_times_groups}
\end{table}

While doing our analysis we also observed that an important dimension of the 
``quickness'' of the users relates to how often they share videos on Twitter. 
Figure~\ref{fig:videoCatTime} shows the median per-user median of the 
inter-event times for users divided into deciles according to the number of 
YouTube videos they have shared. The inter-event times are given in hours and 
range from 352 hours for the least active to 38 hours for the most active users. 
As the difference is quite striking, we inspected term clouds for the Twitter 
bios of the least active YouTube sharers and  the most active YouTube sharers. 
Interestingly, the two are quite similar, apart from a prominent ``YouTube'' for 
the most active users, indicating that the difference in lag time is related to 
the activity level, not topical interests.

\subsection{What is Shared Faster in Twitter}
\begin{figure}[t]
\centering
\epsfig{file=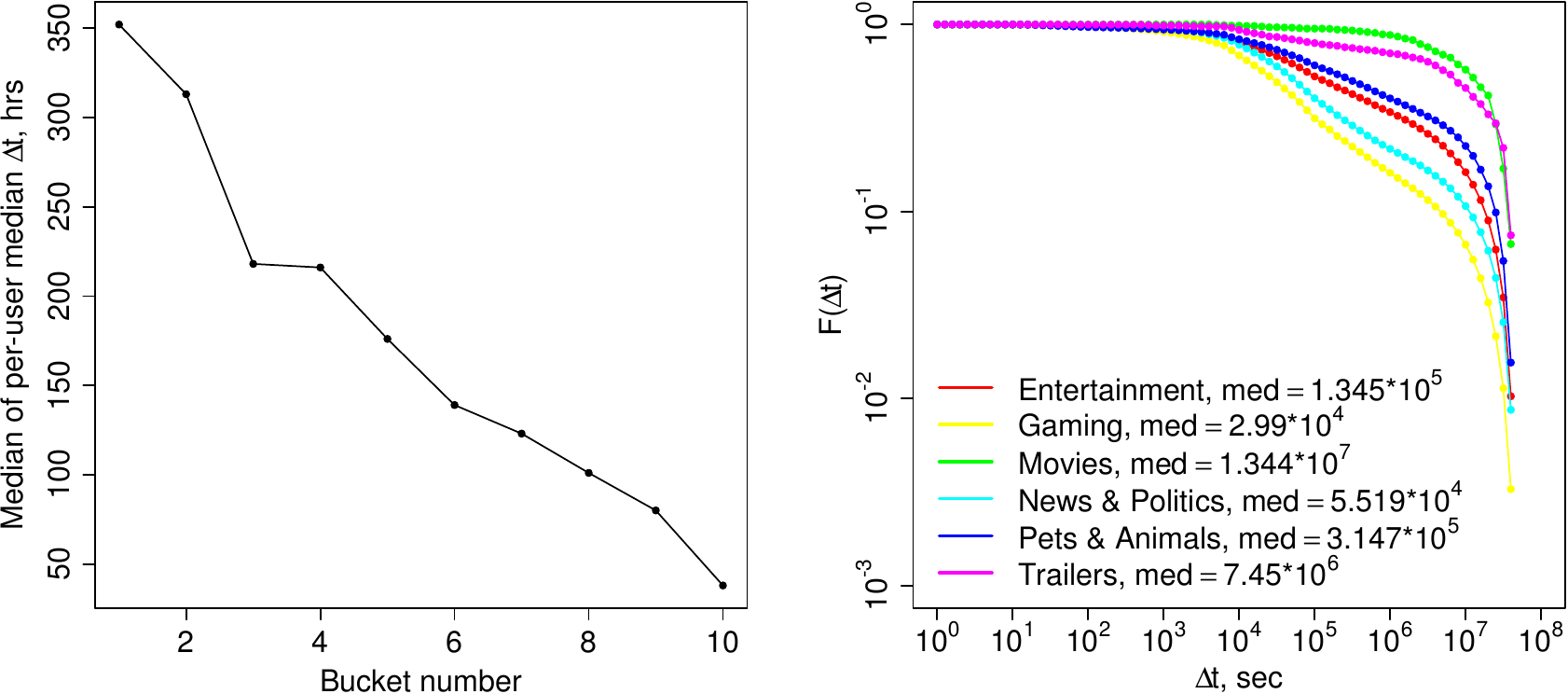, width=\textwidth}
 \caption{Median of per-user median inter-event times for users bucketed (into 
deciles) by the number of YouTube videos shared (left). Accumulative time 
distribution $F(\Delta t)$ of videos belonging to various YouTube categories 
(right).}
\label{fig:videoCatTime}
\end{figure}

To answer question b) we performed system inter-event time analysis and 
distributed time lags in $T$ into relevant video category. If $T_{C}$ is a 
collection of system lags of set of videos belonging to category $C$ 
($\text{\emph{VIDEOS}}^{C}$), then $\Delta t^{w}_{v} \in T_{C}, \text{\emph{if 
}} v \in \text{\emph{VIDEOS}}^{C}$. YouTube provides 19 video categories, in 
Figure~\ref{fig:videoCatTime} inter-event time of 6 categories which exhibit 
different patterns time distribution are shown, due to space limits. Remaining 
13 video categories lag show similar patterns as \emph{Entertainment} and 
\emph{Pets \& Animals}. Our findings show that among all videos, \emph{Gaming} 
and \emph{News \& Politics} videos are the fastest shared with median time of 8 
and 15 hours respectively, \emph{Movies} and \emph{Trailers} have the greatest 
lag between video uploaded and being tweeted with median of 5 and 3 months 
respectively.

\section{Video Popularity Analysis}\label{sec:videoAnalysis}

\subsection{Forecasting Video Popularity}

In this section, we present our work on early indicators of the
popularity of a video, i.e., its amount of views a sufficient amount
of time after its creation.  Our approach is based on analyzing the
Twitter attention to the video in the first moments after its
creation, including the user profile information explained above.  For
this task, we filter our data following the cutoff date explained in
Section \ref{sec:earlyadopters}, and restrict our analysis to videos
that were created before June 1st 2013, a total of 4,822,675 videos
created more than a month before the data retrieval date. We estimate
the popularity of a video through the amount of views more than a
month after its creation, following previous approaches by Szabo and Huberman 
\cite{Szabo2010}, in line with the very fast decay of views that most
videos have in YouTube as shown in Crane and Sornette \cite{Cranesor}.

For each video, we analyze its Twitter attention during the first week
after its creation.  We remove from our analysis all videos that,
during this first week, did not have any sharing event in our
data. This removes old videos that were created before Twitter grew to
its actual user base, leaving us with a set of 276,488 videos.  To
analyze the role of user interests and promotions, we divide our
analysis of Twitter data in two subsets: one only based on promotional
users, and one based on non-promotional users.  After such filtering,
we have a total amount of 1,200,924 shares and 182,135 videos from
promotional users, and 779,821 shares and 133,373 videos from
non-promotional users. Note that these two datasets are disjoint in
terms of Twitter data. No Twitter share is taken into account in
both, but they overlap in 17,093 videos.

\subsection{Twitter Video Metrics}

We measure the early Twitter attention towards a video aggregating two
types of data: i) amount of tweets or attention volume, and ii)
reputation metrics calculated from the follower network and retweeting
behavior of the users involved.  For each video, we computed five
metrics of Twitter attention that summarize different factors that
potentially increase video popularity:

\begin{table} [b]
\centering
\begin{tabular} {|c|c|c|}
  \hline
  Amount of shares   &    Exposure    & Social impact\\
  $S_v = \sum\limits_{u \in U_v} n_v(u)$  &   $E_v = \sum\limits_{u \in
    U_v} f(u)$  & $I_v = \sum\limits_{u \in U_v} R_0(u)$ \\\hline
  Second-order exposure    &    \multicolumn{2}{|c|}{Share of voice} \\
  $\mathcal{E}_v = \sum\limits_{u \in U_v} \sum\limits_{u' \in F(u)}
  f(u')$ &  \multicolumn{2}{|c|}{
    $ A_v = \sum\limits_{u \in U_v}   f(u) / \langle
    f^{-1}(u')\rangle_{u' \in F(u)}$ } \\ \hline
\end{tabular}
\caption{\label{table:metrics} Twitter social metrics used related to video 
popularity.}
\end{table}

We measure the total attention in Twitter to a video through the
\emph{amount of shares} $S_v$ during the first week, which were produced
by the set of users that shared the video in the first week, noted as
$U_v \in U$. Each user $u$ created $n_v(u)$ shares of the video, which
were received by the set of followers of those users. We define the
\emph{exposure} $E_v$ of a video as the sum of followers of the users that
shared the video in the first week, where $F(u)$ is the set of
followers of user $u$, and $f(u) = |F(u)|$. This measure approximates
the size of the first order neighborhood of the accounts sharing the
video, overcounting their common friends.

We aggregate the \emph{social impact} $I_v$ of the users that shared
the video estimated as their mean amount of retweets for tweets
with nonzero retweets ($R_0(u)$). To improve estimation of the reputation of
the users sharing the video in the first week, we approximate the size
of the second-order neighborhood of the users that shared the
video. For this, we calculate the \emph{second-order exposure}
$\mathcal{E}_v$, as the sum of the amount of followers of the
followers of the users that shared the video.

Each user exposed to the shares of the video is subject to have its
attention diluted over a set of different information sources. For
this reason, we calculate the \emph{share of voice} $A_v$ of the early
users, as the ratio of their amount of followers divided by the
average amount of users followed by their followers, where $f^{-1}(v)$
is the amount of users that $v$ follows. This way, we correct the case
of users with many followers, who would give a lower share of voice if
they follow a large amount of other users. On the other hand, a user
with a low amount of followers can have a large share of voice, when
its followers do not follow many other accounts.

We use these five metrics to create a video vector with a sixth
dimension being its final amount of views. In the following, we
present our analysis of the relations between these five metrics and
the popularity of a video.

\subsection{Factors Influencing Video Popularity}

The distribution of views per video, as well as the other metrics
explained above, have large variance and are skewed to the right. To
avoid the uneven leverage of extreme values of these distributions, we
have applied a logarithmic transformation to each one of them, reducing
their variance but keeping their rank. In the first step of our
analysis, we computed correlation coefficients
between the logarithm of the amount of views and the other five
variables. The results for promotional and non-promotional data are
summarized in Table \ref{video-cors}, revealing significant
correlations for all of them. Some of this correlations are of very
low magnitude or even negative sign, suggesting a more careful
analysis.

\begin{table} [b]
\centering
 \begin{tabular} 
{|c|@{\hspace{0.25em}}c@{\hspace{0.25em}}@{\hspace{0.25em}}c@{\hspace{0.25em}}|@
{\hspace{0.5em}}c@{\hspace{0.5em}}@{\hspace{0.5em}}c@{\hspace{0.5em}}|@{\hspace{
0.5em}}c@{\hspace{0.5em}}@{\hspace{0.5em}}c@{\hspace{0.5em}}|@{\hspace{0.5em}}c@
{\hspace{0.5em}}@{\hspace{0.5em}}c@{\hspace{0.5em}}|@{\hspace{0.5em}}c@{\hspace{
0.5em}}@{\hspace{0.5em}}c@{\hspace{0.5em}}|}
 \hline
 & \multicolumn{10}{c|}{$X$}\\ \hline
 \scriptsize Type & \multicolumn{2}{c}{$S_v$} & \multicolumn{2}{c}{$E_v$} &  
\multicolumn{2}{c}{$I_v$} & \multicolumn{2}{c}{$\mathcal{E}_v$}    &  
\multicolumn{2}{c|}{$A_v$}    \\ \hline
 \scriptsize Nonpr  & \scriptsize 0.3 & \scriptsize  0.3 &  \scriptsize  0.1  & 
\scriptsize  0.08 &  \scriptsize  0.4  & \scriptsize 0.4 &  \scriptsize 0.27  & 
\scriptsize 0.29 & \scriptsize -0.05 & \scriptsize -0.16 \\
\hline
\scriptsize Promo  & \scriptsize 0.18 & \scriptsize  0.19 &  \scriptsize  0.16  
& \scriptsize  0.16 &  \scriptsize  0.28  & \scriptsize 0.26 &  \scriptsize 0.13 
& \scriptsize 0.09 & \scriptsize 0.08  & \scriptsize 0.04\\
\hline
\end{tabular}
\caption{\label{video-cors}Pearson's (first value) and Spearman's (second value) 
correlation coefficients
  between video views and Twitter measures: $\rho(log(V_v), log(X))$, all with 
$p<10^{-10}$}
\end{table}

\begin{figure}[t]
\centerline{
 \includegraphics[width=0.48\textwidth]{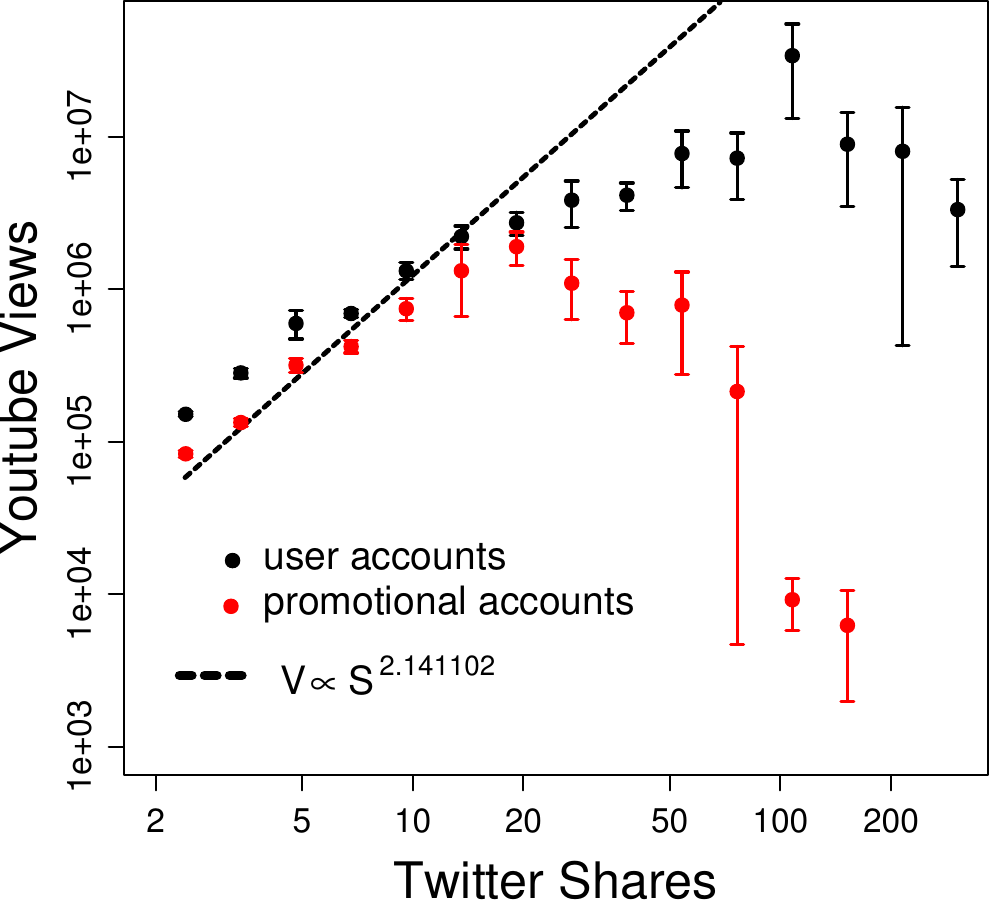} \quad
\includegraphics[width=0.48\textwidth]{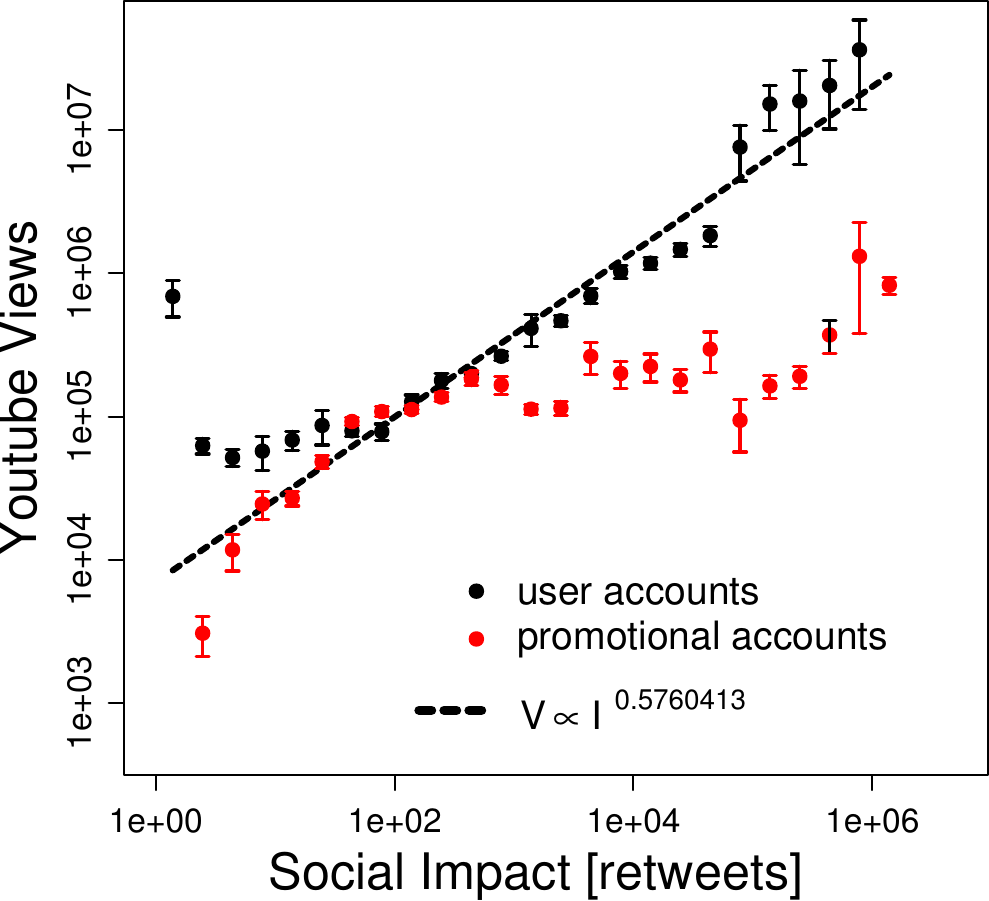}}
\caption{Mean amount of views videos binned by amount of shares (left)
  and social impact of early adopters (right). Error bars show
  standard error, and dashed lines regression results.}
  \label{views-shares-rt}
\end{figure}

Our first observation is that the amount of shares in the first week
of a video is a better predictor for its popularity in the case of
non-promotional users and promotional ones ($\rho=0.184$ vs
$\rho=0.298$).  The left panel of Figure \ref{views-shares-rt} shows
the mean amount of views of videos binned exponentially by their
amount of Twitter shares.

The two types of user activities diverge after 20 shares in the first
week, where for the case of non-promotional users the amount of views
appears to be increasing but saturating. Regression on a power-law
relation between views and shares $V \propto S^\alpha$ reveals a
superlinear scaling with $\alpha=2.18 \pm 0.02$, i.e., the final
views of a video has a quadratic relation to the amount of regular
user shares in the first week.  As an example of this superlinear
growth, the mean amount of views for videos with 2 shares in the first
week is 151,374.5, for videos with 7 shares is 644,522.4, and for
videos with 12 shares is 2,349,317. This gives an increase of almost
500K views for the five shares after the first two, but an increase of
more than 1.7M for the five shares after the first seven.

The diverging pattern in both types of user activity reveals that,
when promotional accounts share the same video more than 20 times in
the same week, the final amount of views does not increase. In fact,
there is a decreasing pattern of views, suggesting the existence of
information overload or spamming behavior in promotional users.

For both types of Twitter users, the aggregated social impact in terms
of mean retweet rates is the best predictor for the popularity of a
video ($\rho=0.394$ and $\rho=0.28$). The right panel of
Fig.\ \ref{views-shares-rt} shows the mean view values for bins of the
aggregated social impact, with the result of regression of the form
$V \propto I^\beta$, where $\beta=0.576 \pm 0.004$ for
non-promotional users and $\beta=0.358 \pm  0.003$ for
promotional ones. This result reveals a sublinear relation between the
amount of views and the social impact of the accounts that shared the
video in the first week, close to a square root.

\begin{figure}[t]
\centerline{
\includegraphics[width=0.48\textwidth]{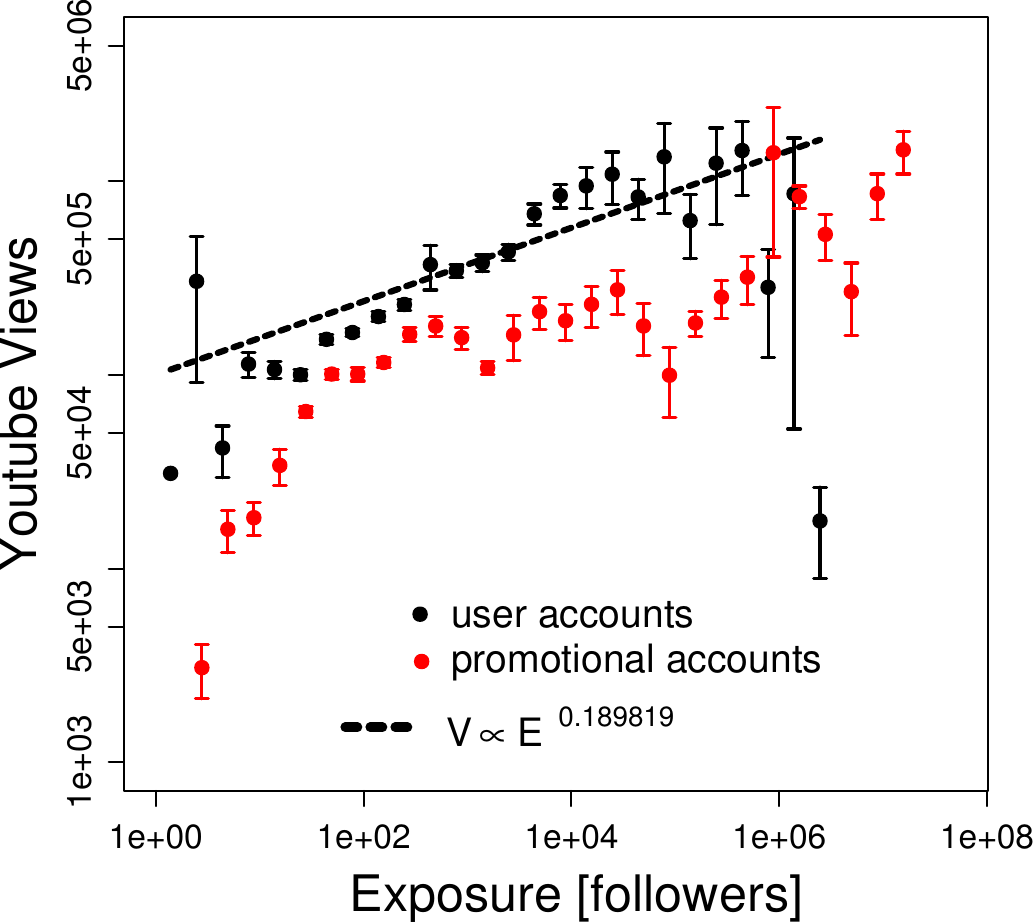}\quad   
\includegraphics[width=0.48\textwidth]{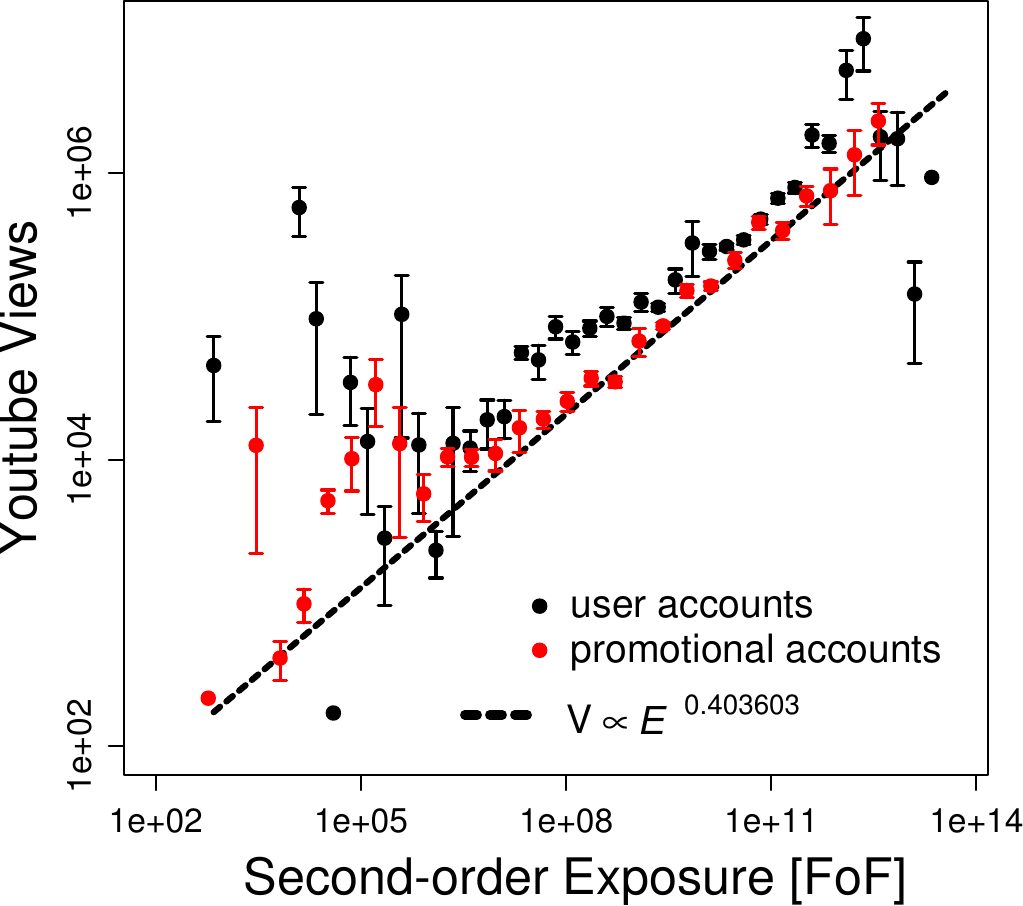}}
\caption{Mean amount of views videos binned by first and second order
  exposure. Error bars show standard error, and dashed lines
  regression results.}
  \label{views-fof}
\end{figure}

The amount of views of videos showed a low positive correlation
coefficient with the exposure of the shares in the first week,
measured through amount of followers. The left panel of
Fig. \ref{views-fof} shows the mean amount of views versus the
exposure in the first week, revealing a very soft increasing pattern
in both. On the other hand, the amount of views has a more substantial
correlation with the second-order exposure, with correlation
coefficients of $0.268$ and $0.126$ for regular and promotional users
respectively. The right panel of Fig. \ref{views-fof} shows this
stronger relation, with a regression result of exponent $0.404 \pm
0.004$ for non-promotional users, and of $0.155 \pm 0.003$ for
promotional ones. This comparison reveals that the second-order
exposure is a much better predictor for the popularity of a video than
the amount of followers of the initial sharers. This result calls for
more stylized reputation metrics that take into account global
information beyond amount of followers and retweet rates, for example
centrality \cite{Gayo-Avello2010}, or coreness \cite{Garcia2013d} metrics.

Finally, the aggregated share of voice of the accounts that shared the
video during the first week did not provide clear results, with a
significant negative correlation of $-0.047$ for non-promotional users,
and of $0.076$ for promotional ones. This suggest that, if information
overload and competition for attention are present in Twitter, they
need to be measured with more precise approximations that the
correction we presented in the previous section. Nevertheless, the
share of voice of the users sharing a video still contains relevant
information that we introduce in the regression model we explain
below.

\subsection{Combining Data in a Regression Model}

The above results show the pairwise relation between the amount of
views of a video and each one of our five Twitter metrics. This
analysis ignores the possible effect of the combination of different
metrics, as it can be expected that they are correlated with each
other.  To provide a deeper analysis on how these Twitter metrics
influence the final amount of views, we propose a substitutes model in
which the products of powers of each variable are proportional to the
final amount of views:

\begin{equation}
V_v \propto \cdot  S_v^\alpha   \cdot   I_v^\beta  \cdot E_v^\gamma  \cdot
\mathcal{E}_v^\delta \cdot A_v^\kappa
\label{eq-model}
\end{equation}

This model is equivalent to a linear regression model after the
logarithmic transformation of all the independent variables. Training
this regressor on the promotional user data gives $R^2=0.107$,
explaining about 10\% of the variance of $log(V)$. On the
non-promotional user dataset, the regressor achieves $R^2=0.199$,
explaining almost 20\% of the variance of the final amount of views of
a video based exclusively on information extracted from Twitter. This
opens the possibility to improve previous models that used only data
from YouTube \cite{Szabo2010, Pinto2013}, which could also be combined with data 
from other online communities, as previously done in Soysa et al. 
\cite{Soysa2013} with a limited sample of Facebook data.

The estimated coefficients for the exponents of Eq. \ref{eq-model} are reported 
in Table \ref{video-reg}, which allow us to compare the size of the effects of 
each Twitter metric. This analysis reveals the lack of relevance of the 
first-order exposure for the case of non-promotional users as also shown in Cha 
et al. \cite{ChaHBG10}. The correlation between first order exposure and views 
shown in Fig. \ref{views-fof} is a confound due to the correlation of exposure 
with other metrics, such as impact or second-order exposure.

\begin{table}[t]
\centering
\begin{tabular} {|c|c|c|c|c|c|}
 \hline\rule{-4pt}{10pt}
 Type & $S_v$ & $E_v$ &  $I_v$ & $\mathcal{E}_v$    &  $A_v$    \\ \hline
 Not promo  &  $1.083^*$  &  $0.096$ &  $0.449^*$ &  $0.118^*$ & $-0.102^*$ \\
 Promotional &  $0.612^*$&  $0.164^*$ &   $0.307^*$ & $0.079^*$  & $0.030$ \\
\hline
\end{tabular}
\caption{\label{video-reg}Regression coefficients for
  Eq. \ref{eq-model}. Significance level $^*$ $p<10^{-9}$, or $p>0.01$
  otherwise.}
\end{table}

To assess the prediction power of our model for non-promotional users,
we transformed the regression problem to a dichotomous classification,
in which we tag a video as \emph{popular} if it gathered at least
10,000 views. Using the regression model explained above, we can
predict if a video will reach more than 10,000 views based on the first
week of Twitter activity. If the estimator of Eq. \ref{eq-model} gives
a value above 10,000, we classify the video as \emph{popular}.

We performed 10-fold cross validation on the non-promotional users
dataset, fitting the regressor to 90\% of the data and validating it
on the rest 10\%. The mean base rate of popular videos for the 10
evaluations is $0.493$, and our predictor achieves a mean precision of
$0.715$ and a mean recall of $0.534$ for the \emph{popular}
class. Both values are significantly above the precision of random
classifiers over the same partitions, which produced a precision of
$0.492$ and a recall of $0.494$.  This experiment shows that, using
Twitter data only, a prediction can achieve a precision value much
higher than expected from a random classifier.

\section{Conclusions}\label{sec:conclusions}
We gathered a high-quality dataset based on the combination of two
sources: 17 million unique public tweets for 87K users on Twitter and YouTube 
data for 5 million videos. Through this combination of data sets, we could 
obtain novel, detailed insights into who watches (and shares) what on YouTube, 
and when (that is, how quickly). We applied a set of heuristics to infer 
demographic data including gender, location, political alignment, and interests.
We designed a new method to distinguish promotional Twitter accounts, who almost 
exclusively share their own YouTube videos and validated our expectation that 
promotional users share their own videos much faster than regular ones. Our 
results also include a new method to characterize different user segments
in terms of YouTube categories, Twitter activity, and Twitter user
bios. These allowed us to analyze the relation between demographic
factors and the features of YouTube videos, including their amount of
views. Our detailed statistical analysis reveals correlations between
Twitter behavior and YouTube video content. In addition, our clustering analysis 
shows that the topic preferences of the two platforms are largely aligned. Our 
results on politics quantitatively show that politically right users are further 
from the center than politically left users, and among all video categories News 
\& Politics correlates most with social and with sharing behavior. Our detailed 
analysis distinguishes which user types share videos earlier/later as well as 
which video classes are shared earlier/later. Finally, we developed a regression 
model for the effect of early Twitter video shares by influential users on the 
final view count and we conclude by observing that second-order neighborhoods 
and retweet rates are much better predictors of ultimate video popularity than 
raw follower counts.

\section{Acknowledgments}
ETH received funding from the Swiss National Science Foundation under the grant 
CR21I1\_146499 / 1. The authors would like to thank GNIP for providing data 
during a free trial period of their Powertrack product.

\bibliographystyle{abbrv}
\bibliography{wsdm611-abisheva_youtube_twitter} 

\end{document}